\documentclass{article}

\newtheorem{proposition}{Proposition}
\newtheorem{theorem}{Theorem}
\newtheorem{corollary}{Corollary}
\newtheorem{lemma}{Lemma}

\begin{document}


\begin{center}
\vskip 1cm{\LARGE\bf 
Searching of gapped repeats and \\
\vskip .15in
subrepetitions in a word}
\vskip 1cm
{\large Roman Kolpakov}\\
Lomonosov Moscow State University \\ 
Leninskie Gory, Moscow, 119992 Russia \\
{\tt foroman@mail.ru}
\vskip .5cm
{\large Mikhail Podolskiy}\\
Lomonosov Moscow State University \\ 
Leninskie Gory, Moscow, 119992 Russia \\
{\tt mpodolskii@inbox.ru}
\vskip .5cm
{\large Mikhail Posypkin}\\
Institute for Information Transmission Problems \\ 
Bolshoy Karetny per., Moscow, 127994 Russia \\
{\tt mposypkin@gmail.com}
\vskip .5cm
{\large Nickolay Khrapov}\\
Institute for Information Transmission Problems \\ 
Bolshoy Karetny per., Moscow, 127994 Russia \\
{\tt nkhrapov@gmail.com}
\end{center}

\vskip .2 in

\begin{abstract}
A gapped repeat is a factor of the form $uvu$ where $u$ and $v$ are
nonempty words. The period of the gapped repeat is defined as $|u|+|v|$.
The gapped repeat is maximal if it cannot be extended to the left or to 
the right by at least one letter with preserving its period. The gapped 
repeat is called $\alpha$-gapped if its period is not greater than $\alpha |v|$.
A $\delta$-subrepetition is a factor which exponent is less than~2 but is not
less than $1+\delta$ (the exponent of the factor is the quotient of the length
and the minimal period of the factor). The $\delta$-subrepetition is maximal if 
it cannot be extended to the left or to the right by at least one letter with 
preserving its minimal period. We reveal a close relation between maximal gapped 
repeats and maximal subrepetitions. Moreover, we show that in a word of length~$n$
the number of maximal $\alpha$-gapped repeats is bounded by $O(\alpha^2n)$ and
the number of maximal $\delta$-subrepetitions is bounded by $O(n/\delta^2)$.
Using the obtained upper bounds, we propose algorithms for finding all maximal 
$\alpha$-gapped repeats and all maximal $\delta$-subrepetitions in a word of length~$n$.
The algorithm for finding all maximal $\alpha$-gapped repeats has $O(\alpha^2n)$ time 
complexity for the case of constant alphabet size and $O(n\log n + \alpha^2n)$ time 
complexity for the general case. For finding all maximal $\delta$-subrepetitions we
propose two algorithms. The first algorithm has $O(\frac{n\log\log n}{\delta^2})$ time 
complexity for the case of constant alphabet size and $O(n\log n +\frac{n\log\log n}{\delta^2})$ 
time complexity for the general case. The second algorithm has 
$O(n\log n+\frac{n}{\delta^2}\log \frac{1}{\delta})$ expected time complexity.
\end{abstract}

\section{Inroduction}

Let $w=w[1]w[2]\ldots w[n]$ be an arbitrary word. The length of~$w$ is 
denoted by $|w|$. A fragment $w[i]\cdots w[j]$ of~$w$, where $1\le i\le j\le n$, 
is called a {\it factor} of~$w$ and is denoted by $w[i..j]$.
Note that for factors we have two different notions of equality:
factors can be equal as the same fragment of the original word or as the 
same word. To avoid this ambiguity, we will use two different notations:
if two factors $u$ and $v$ are the same word (the same fragment of the original 
word) we will write $u=v$ ($u\equiv v$). For any $i=1,\ldots,n$ the factor 
$w[1..i]$ ($w[i..n]$) is called a {\it prefix} (a {\it suffix}) of~$w$.
By positions in~$w$ we mean the order numbers $1, 2,\ldots ,n$ of letters
of the word~$w$. For any factor~$v\equiv w[i..j]$ of~$w$ the positions $i$ 
and $j$ are called {\it start position} of~$v$ and {\it end position} of~$v$
and denoted by ${\rm beg} (v)$ and ${\rm end} (v)$ respectively. The factor~$v$
{\it covers} a letter $w[k]$ if ${\rm beg} (v)\le k\le {\rm end} (v)$.
For any two factors $u$, $v$ of~$w$ the factor $u$ {\it is contained}
({\it is strictly contained}) in~$v$ if ${\rm beg} (v)\le {\rm beg} (u)$
and ${\rm end} (u)\le {\rm end} (v)$ (if ${\rm beg} (v)<{\rm beg} (u)$
and ${\rm end} (u)<{\rm end} (v)$). Let $u$, $v$ be two factors of~$w$
such that ${\rm beg} (v)={\rm end} (u)+1$. In this case we say that
$v$ {\it follows}~$u$. The number ${\rm end} (u)$ is called {\it the frontier} 
between the factors $u$ and $v$. A factor~$v$ {\it contains} a frontier~$j$
if ${\rm beg} (v)-1\le j\le {\rm end} (v)$. If some word $u$ is equal to 
a factor~$v$ of~$w$ then $v$ is called {\it an occurence} of~$u$ in~$w$.

A positive integer $p$ is called  a {\it period} of~$w$ if $w[i]=w[i+p]$ 
for each $i=1,\ldots ,n-p$. We denote by $p(w)$ the minimal period of~$w$
and by $e(w)$ the ratio $|w|/p(w)$ which is called the {\it exponent} of~$w$.
A word is called {\it primitive} if its exponent is not an integer greater
than~1. By repetition in a word we mean any factor of exponent greater than 
or equal to~2. Repetitions are fundamental objects, due to their primary importance 
in word combinatorics~\cite{Lothaire83} as well as in various applications, such as 
string matching algorithms~\cite{GaliSeiferas83,CrochRytter95}, molecular 
biology~\cite{Gusfield97}, or text compression~\cite{Storer88}.
The simplest and best known example of repetitions is factors of the form $uu$, 
where $u$ is a nonempty word. Such repetitions are called {\it squares}.
We call the first (second) factor~$u$ of the square $uu$ {\it the left (right)
root} of this square. Avoiding ambiguity\footnote{Note that the period of a square 
is not necessarily the minimal period of this word.}, by {\it the period} of a square 
we mean the length of its roots. A square is called {\it primitive} if its roots
are primitive. The questions concerned to squares are well studied in the literature.
In particular, it is known (see, e.g.,~\cite{CrochRytter95}) that a word of length~$n$
contains no more than $\log_{\varphi} n$ primitive squares. In~\cite{Crochemor81}
an $O(n\log n)$-time algorithm for finding of all primitive squares in a word of length~$n$
is proposed. In~\cite{GusfStoye04} an algorithm for finding of all primitive squares 
in a word of length~$n$ with time complexity $O(n+S)$ where $S$ is the size of output
is proposed for the case of constant alphabet size.

A repetition in a word is called {\it maximal} if this repetition cannot be extended 
to the left or to the right in the word by at least one letter with preserving its minimal period.
More precisely, a repetition $r\equiv w[i..j]$ in~$w$ is called {\it maximal}
if it satisfies the following conditions:
\begin{enumerate}
\item if $i>1$, then $w[i-1]\neq w[i-1+p(r)]$,
\item if $j<n$, then $w[j+1-p(r)]\neq w[j+1]$.
\end{enumerate}
Maximal repetitions are usually called {\it runs} in the literature. Since runs contain 
all the other repetitions in a word, the set of all runs can be considered as a compact 
encoding of all repetitions in the word which has many useful applications (see, 
for example,~\cite{Crochetal1}). For any word~$w$ we will denote by ${\cal R} (w)$  
the set of all maximal repetitions in~$w$ and by ${\rm E}(w)$ the sum of exponents
of all maximal repetitions in~$w$. The following facts are proved in~\cite{KK00}.

\begin{theorem}
${\rm E}(w)=O(n)$ for any~$w$.
\label{sumexp}
\end{theorem}

\begin{corollary}
$|{\cal R}(w)|=O(n)$ for any~$w$.
\label{onmaxrun}
\end{corollary}

Moreover, in~\cite{KK00} an $O(n)$ time algorithm for finding of all runs in a word of 
length~$n$ is proposed for the case of constant alphabet size (in the case of arbitrary
alphabet size all runs in a word of length~$n$ can be found in $O(n\log n)$ time). 
Further many papers were devoted to obtaining more precise upper bounds on ${\rm E}(w)$ 
and $|{\cal R}(w)|$. In our knowledge, at present time the best upper bounds for these 
values are obtained in~\cite{CrochIlieTinta} and~\cite{Crochetal11}.

A natural generalization of squares is factors of the form $uvu$ where $u$ and $v$ are
nonempty words. We call such factors {\it gapped repeats}. In the gapped repeat $uvu$
the first (second) factor~$u$ is called {\it the left (right) copy}, and $v$ is called 
{\it the gap}. By {\it the period} of this gapped repeat we will mean the value $|u|+|v|$.
For a gapped repeat~$\sigma$ we denote the length of copies of~$\sigma$ by $c(\sigma)$
and the period of~$\sigma$ by $p(\sigma)$. By $(u', u'')$ we will denote the gapped repeat 
with the left copy~$u'$ and the right copy~$u''$.  Note that gapped repeats with distinct 
periods can be the same factor, i.e. can have the same both start and end positions
in the word. In this case, for convenience, we will consider this repeats as different ones,
i.e. a gapped repeat is not determined uniquely by its start and end positions in the word
because this information is not sufficient for determining the both copies and the gap of
the repeat. For any real $\alpha>1$ a gapped repeat~$\sigma$ is called {\it $\alpha$-gapped} 
if $p(\sigma)\le\alpha c(\sigma)$. Analogously to repetitions, we can introduce the notion 
of maximality for gapped repeats. A gapped repeat $(w[i'..j'], w[i''..j''])$ in~$w$ is called 
{\it maximal} if it satisfies the following conditions:
\begin{enumerate}
\item if $i'>1$, then $w[i'-1]\neq w[i''-1]$,
\item if $j''<n$, then $w[j'+1]\neq w[j''+1]$.
\end{enumerate}
In other words, a gapped repeat in a word is maximal if its copies cannot be extended 
to the left or to the right in the word by at least one letter with preserving its period.
Note that any $\alpha$-gapped repeat is contained either in a determined uniquely maximal
$\alpha$-gapped repeat with the same period or, otherwise, in a determined uniquely maximal 
repetiton which minimal period is a divisor of the period of the repeat. Therefore, for
computing all $\alpha$-gapped repeats in a given word it is enough to find all maximal 
$\alpha$-gapped repeats and all maximal repetitions in this word. Thus, taking into account
the existence effective algorithms for finding of all runs in a word, we can conclude that
the problem of computing all $\alpha$-gapped repeats in a word is reduced to the problem
of finding all maximal $\alpha$-gapped repeats in a word. The set of all maximal $\alpha$-gapped 
repeats in~$w$ will be denoted by ${\cal GR}_{\alpha}(w)$. The problem of finding gapped repeats
in a word was investigated before. In particular, it is shown in~\cite{Brodal00} that all maximal 
gapped repeats with a gap length belonging to a specified interval can be found in a word  
of length~$n$ with time complexity $O(n\log n+S)$ where $S$ is the size of output.
An algorithm for finding in a word all gapped repeats with a fixed gap length is proposed 
in~\cite{KK00a}. The proposed algorithm has time complexity $O(n\log d+S)$ where $d$ is the gap 
length, $n$ is the word length, and  $S$ is the size of output.

Another natural generalization of repetitions is factors with exponents strictly less than~2.
We will call such factors {\it subrepetitions}. More precisely, for any~$\delta$ such that
$0<\delta<1$ by $\delta$-subrepetition we mean a factor~$v$ such that $1+\delta\le e(v)<2$.
Note that the notion of maximal repetition is directly generalized to the case of subrepetitions:
maximal subrepetitions are defined exactly in the same way as maximal repetitions.
Further we reveal a close relation between maximal subrepetitions and maximal gapped repeats.
Some results concerning the possible number of maximal subrepetitions in words were
obtained in~\cite{KKOch}. In particular, it was proved that the number of maximal 
$\delta$-subrepetitions in a word of length~$n$ is bouned by $O(\frac{n}{\delta}\log n)$.

The aim of our research is to develop effective algorithms of finding maximal gapped repeats
and maximal subrepetitions in a given word. Firstly we estimate the number of maximal $\alpha$-gapped 
repeats in a word of length~$n$. In the paper we prove $O(\alpha^2n)$ upper bound on this number.
From this bound we derive $O(n/\delta^2)$ upper bound on the number of maximal $\delta$-subrepetitions
in a word of length~$n$. Using the obtained bound on the number of maximal gapped repeats in a word,
we show that in the case of constant alphabet size all maximal $\alpha$-gapped repeats in a word of length~$n$ 
can be found in $O(\alpha^2n)$ time. For finding all  maximal $\delta$-subrepetitions in the word
we propose two algorithms. The first algorithm has time complexity $O(\frac{n\log\log n}{\delta^2})$ 
in the case of constant alphabet size and $O(n\log n+\frac{n\log\log n}{\delta^2})$ in the general case. 
The second algorithm has $O(n\log n+\frac{n}{\delta^2}\log\frac{1}{\delta})$ expected time 
complexity. 

\section{Auxiliary definitions and results}

Further we will consider an arbitrary word $w=w[1]w[2]\ldots w[n]$ of length~$n$.
Recall that any repetition~$r$ in~$w$ is extended to just one maximal repetition $r'$
with the same minimal period. We will call the repetition $r'$ {\it the extension} of~$r$.
We will use the following quite evident fact on maximal repetitions (see, e.g.,~\cite{Lothaire05}[Lemma 8.1.3]).

\begin{lemma}
Two distinct maximal repetitions with the same minimal period~$p$ can not
have an overlap of length greater than or equal to~$p$.
\label{overlap}
\end{lemma}

For primitive words the following well-known fact takes place (see, e.g.,~\cite{algonstr}).

\begin{lemma}[primitivity lemma]
If $u$ is a primitive word, then $u$ can not be strictly contained in the square $uu$.
\label{onprimit}
\end{lemma}

Using Lemma~\ref{onprimit}, it is easy to prove

\begin{proposition}
If a square $uu$ is primitive, for any two distinct occurrences $v'$
and $v''$ of $uu$ in~$w$ the inequality $|{\rm beg} (v')-{\rm beg} (v'')|\ge |u|$
holds.
\label{onprimsqr}
\end{proposition}

\begin{corollary}
If a square $uu$ is primitive, any factor~$v$ contains no more than $|v|/|u|$
occurrences of $uu$.
\label{numcycsqr}
\end{corollary}

Let $r$ be a repetition in the word~$w$. We call any factor 
of~$w$ which has the length $p(r)$ and is contained in~$r$ a 
{\it cyclic root} of~$r$. The cyclic root which is the prefix
(suffix) of~$r$ is called {\it prefix (suffix)} cyclic root of~$r$.
Note that for any cyclic root $u$ of~$r$ the word~$r$ is a factor of 
the word $u^k$ for some big enough~$k$. So it follows from 
the minimality of the period $p(r)$ that any cyclic root of~$r$ 
has to be a primitive word. Hence any two adjacent cyclic roots 
of~$r$ form a primitive square with the period $p(r)$ 
which is called a {\it cyclic square} of~$r$. The cyclic square 
which is the prefix (suffix) of~$r$ is called {\it prefix (suffix)} 
cyclic square of~$r$. The following proposition can be easily
obtained from Lemma~\ref{onprimit}.

\begin{proposition}
Two cyclic root $u'$, $u''$ of a repetition~$r$ are equal if and only if
${\rm beg} (u')\equiv {\rm beg} (u'')\pmod{p}$.
\label{oncycroot}
\end{proposition}

Thus we have

\begin{corollary}
Any repetition~$r$ contains no more than $|r|/p(r)$ equal cyclic roots.
\label{numcycroot}
\end{corollary}

For obtaining our results, we introduce the following classification of maximal gapped
repeats. We say that a maximal gapped repeat is {\it periodic} if the copies
of this repeat are repetitions. The set of all periodic maximal $\alpha$-gapped repeats
in the word~$w$ is denoted by ${\cal PP}_{\alpha}$. A gapped maximal repeat is
called {\it prefix} ({\it suffix}) {\it semiperiodic} if the copies of this
repeat are not repetitions, but these copies have a prefix (suffix) satisfying
the following conditions:
\begin{enumerate}
\item this prefix (suffix) is a repetition;
\item the length of this prefix (suffix) is not less than the half of the copies length.
\end{enumerate}
In a copy of a prefix semiperiodic repeat the longest prefix satisfying the above
conditions is called {\it periodic prefix} of this copy. The periodic prefixes
of the copies of a prefix semiperiodic repeat are also called periodic prefixes
of this repeat. The set of all prefix (suffix) semiperiodic $\alpha$-gapped maximal
repeats in the word~$w$ is denoted by~${\cal PSP}_{\alpha}$ (${\cal SSP}_{\alpha}$).
A gapped maximal repeat is called {\it semiperiodic} if it is either prefix or suffix
semiperiodic. The set of all semiperiodic $\alpha$-gapped maximal repeats in the word~$w$ 
is denoted by~${\cal SP}_{\alpha}$. Gapped maximal repeats which are neither periodic
nor semiperiodic are called {\it ordinary}. The set of all ordinary $\alpha$-gapped 
maximal repeats in the word~$w$ is denoted by~${\cal OP}_{\alpha}$.

Let $\delta<1$ and $r$ be a maximal $\delta$-subrepetition in~$w$. Then we can consider 
in~$w$ the repeat $\sigma\equiv (w[{\rm beg} (r)..{\rm end} (r)-p(r)], 
w[{\rm beg} (r)+p(r).. {\rm end} (r)])$. It follows from $e(r)<2$ that $\sigma$ is
gapped. Moreover, $p(\sigma)=p(r)$ and, since $r$ is maximal, it is obvious that 
$\sigma$ is maximal. Since $r$ is a $\delta$-subrepetition, we have also that
$|r|-p(r)\ge\delta p(r)$, so $c(\sigma)=|r|-p(r)\ge \delta p(r)=\delta p(\sigma)$,
i.e. $p(\sigma)\le \frac{1}{\delta}c(\sigma)$. Thus, $\sigma$ is a maximal 
$\frac{1}{\delta}$-gapped repeat in~$w$. We will call the subrepetition~$r$ and 
the repeat~$\sigma$ {\it respective} to each other. Note that for each maximal 
$\delta$-subrepetition~$r$ there exists a maximal $\frac{1}{\delta}$-gapped 
repeat~$\sigma$ respective to~$r$. Moreover, the subrepetition~$r$ is determined 
uniquely by the repeat~$\sigma$, so the same repeat can not be respective to different 
subrepetitions. Thus we have

\begin{proposition}
Let $0<\delta <1$. Then in any word the number of maximal $\delta$-subrepetitions
is no more then the number of maximal $1/\delta$-gapped repeats.
\label{relforep}
\end{proposition}

On the other hand, it is easy to see that a maximal gapped repeat can have no 
a respective maximal subrepetition. Maximal gapped repeats which have respective 
maximal subrepetitions will be called {\it principal}. Thus we have the one-to-one 
correspondence between maximal $\delta$-subrepetitions and principal $\frac{1}{\delta}$-gapped 
repeats in a word. It is easy to check the following fact.

\begin{proposition}
A maximal gapped repeat~$\sigma$ in~$w$ is principal if and only if $p(w[{\rm beg} (\sigma)..{\rm end} (\sigma)])=
p(\sigma)$.
\label{princrit1}
\end{proposition}

Let $\sigma$ be a maximal gapped repeat, and $r$ be a maximal repetition or subrepetition.
We will say that $\sigma$ {\it is stretched} by~$r$ if $\sigma$ is contained in~$r$ and $p(r)<p(\sigma)$
and call $\sigma$ {\it stretchable} if $\sigma$ is stretched by some maximal repetition or subrepetition.
It follows from Proposition~\ref{princrit1} that $\sigma$ is not principal if and only if 
$p(w[{\rm beg} (\sigma)..{\rm end} (\sigma)])<p(\sigma)$, i.e. $\sigma$ is contained in
some maximal repetition or subrepetition with minimal period less than $p(\sigma)$.
So we obtain

\begin{proposition}
A maximal gapped repeat is principal if and only if it is not stretchable.
\label{princrit2}
\end{proposition}

We will say that a gapped repeat~$\sigma$ {\it is stretched} by a gapped repeat~$\sigma'$
if $\sigma$ is contained in~$\sigma'$ and $p(\sigma')<p(\sigma)$. It is easy to see
that a gapped repeat is stretched by a subrepetition if and only if this repeat is
stretched by the gapped repeat respective to this subrepetition. Using this observation,
we can derive the following

\begin{proposition}
A maximal $\delta$-gapped repeat is stretchable if and only if it is stretched by either a maximal repetition
or a maximal $\delta$-gapped repeat.
\label{strcrit}
\end{proposition}

\section{Estimation of the number of maximal repeats and repetitions}

In this section we estimate the number of maximal $\alpha$-gapped repeats in a word.
For convenience sake we assume that $\alpha$ is integer although our proof
can be easily generalized to the case of any~$\alpha$. More precisely, we
prove that for any integer $k\ge 2$ the number of maximal $k$-gapped repeats
in the word~$w$ is $O(nk^2)$. To obtain this bound, we estimate separately
the numbers of periodic, semiperiodic and ordinary maximal $k$-gapped repeats in~$w$.

First we estimate the number of periodic maximal $k$-gapped repeats in~$w$. Let $\sigma =(v', v'')$ be 
an arbitrary repeat from~${\cal PP}_k$. Then the both copies $v'$, $v''$ of $\sigma$ are repetitions
in~$w$ which are extended respectively to some maximal repetitions $r'$, $r''$ with the same minimal
period in~$w$. If $r'$ and $r''$ are the same repetition~$r$ then we call $\sigma$ {\it private} repeat 
and we say that $\sigma$ is {\it generated} by~$r$ . Othervise $\sigma$ is called {\it non-private}.
To estimate the number of private maximal $k$-gapped repeats in~$w$, we use

\begin{lemma}
Any maximal repetition~$r$ generates no more than $e(r)/2$ different private gapped 
maximal repeats.
\label{onpriv}
\end{lemma}

{\bf Proof.} 
Let $r$ be a maximal repetition in~$w$ with the minimal period~$p$, and
$\sigma\equiv (v', v'')$ be a private maximal gapped repeat generated by~$r$.
Denote by $u'$ and $u''$ the prefixes of length~$p$ in $v'$ and $v''$ respectively.
Note that $u'$ and $u''$ are equal cyclic roots of~$r$, so by Proposition~\ref{oncycroot}
we have ${\rm beg} (u')\equiv {\rm beg} (u'')\pmod{p}$. Thus ${\rm beg}(v')\equiv 
{\rm beg} (v'')\pmod{p}$. Therefore, if ${\rm beg} (v')>{\rm beg} (r)$ then 
$w[{\rm beg} (v')-1]=w[{\rm beg} (v'')-1]$ which contradicts that $\sigma$ is maximal.
Thus ${\rm beg} (v')={\rm beg} (r)$, i.e. $v'$ is a prefix of~$r$ and $u'$ is the
prefix cyclic root of~$r$. Similarly we can prove that $v''$ is a suffix of~$r$.
Thus $\sigma$ is determined uniquely by the cyclic root $u''$ which is equal to
the prefix cyclic root of~$r$. Moreover, since $\sigma$ is gapped, $u''$ has to
be contained in the suffix of length $\lfloor |r|/2\rfloor$ in~$r$. By Corollary~\ref{numcycroot}
there exist no more than $|r|/2p=e(r)/2$ cyclic roots satisfying the above conditions
for $u''$. Thus there exist no more than $e(r)/2$ private maximal gapped repeats 
generated by~$r$.

Lemma~\ref{onpriv} implies immediately that the number of private maximal gapped repeats in~$w$
is not greater than ${\rm E}(w)/2$. Thus, taking into account Theorem~\ref{sumexp}, we
obtain

\begin{corollary}
The number of private maximal gapped repeats in~$w$ is $O(n)$.
\label{numpriv}
\end{corollary}

Now let $\sigma$ be non-private, i.e. $r'$ and $r''$ be different maximal repetitions.
Then we choose from the repetitions $r'$ and $r''$ the shortest repetition (if $|r'|=|r''|$
we choose any of these repetitions) and say that $\sigma$ is {\it generated} by the
choosen repetition. More precisely, if the chosen repetition is $r'$ ($r''$) we will 
say that $\sigma$ is generated  {\it from the left} ({\it from the right}) by the 
repetition $r'$ ($r''$). We prove the following fact.

\begin{lemma}
For any maximal repetition~$r$ the number of non-private maximal $k$-gapped repeats
generated by~$r$ is $O(k e(r))$.
\label{numnonpriv}
\end{lemma}

{\bf Proof.}
Let $r$ be an arbitrary maximal repetition with the minimal period~$p$ in~$w$.
We will prove that the number of non-private maximal $k$-gapped repeats
generated from the left by~$r$ is $O(k e(r))$. Since the number of non-private 
maximal $k$-gapped repeats generated from the right by~$r$ can be estimated
similary, it will imply the statement of the lemma. Denote by~$P(r)$ the set of 
all non-private maximal $k$-gapped repeats generated from the left by~$r$.
Let $\sigma\equiv (v, v')$ be an arbitrary repeat from~$P(r)$. Denote by~$r'$ 
the extension of~$v'$ which is the maximal repetition with the same minimal 
period~$p$. If ${\rm beg} (v)>{\rm beg} (r)$ and ${\rm beg} (v')>{\rm beg} (r')$ 
then
$$
w[{\rm beg} (v)-1]=w[p+{\rm beg} (v)-1]=v[p]=v'[p]=w[p+{\rm beg} (v')-1]=w[{\rm beg} (v')-1]
$$
which contradicts that $\sigma$ is maximal. Thus we have either ${\rm beg} (v)={\rm beg} (r)$
or ${\rm beg} (v')={\rm beg} (r')$. We can prove similary that either ${\rm end} (v)={\rm end} (r)$
or ${\rm end} (v')={\rm end} (r')$. Thus we can consider the following four possible cases.

1. ${\rm beg} (v)={\rm beg} (r)$ and ${\rm end} (v)={\rm end} (r)$;

2. ${\rm beg} (v)={\rm beg} (r)$ and ${\rm end} (v')={\rm end} (r')$;

3. ${\rm beg} (v')={\rm beg} (r')$ and ${\rm end} (v)={\rm end} (r)$;

4. ${\rm beg} (v')={\rm beg} (r')$ and ${\rm end} (v')={\rm end} (r')$.

Note that in the case~4 we have $|r'|=|v'|=|v|\le |r|$. Therefore, since $|r|\le |r'|$
by the definition of generated repeat, in this case we obtain that $|r|=|r'|=|v|$, i.e.
${\rm beg} (v)={\rm beg} (r)$ and ${\rm end} (v)={\rm end} (r)$. So the case~4 is actually
a subcase of the case~1. Thus $P(r)=P_1(r)\cup P_2(r)\cup P_3(r)$ where $P_i(r)$ the set of 
all repeats from~$P(r)$ which satify the case~$i$. We will estimate separately $|P_1(r)|$, $|P_2(r)|$, 
and $|P_3(r)|$.

Let $\sigma\in P_1(r)$, i.e. $v\equiv r$. Denote by $u$ and $u'$ the prefixes of length $2p$
in $v$ and $v'$ respectively. Note that in this case $\sigma$ is determined uniquely by
the factor~$u'$. Note also that $u=u'$ and $u$ is the prefix cyclic square of~$r$.
Thus $u'$ is a primitive square with period~$p$ which is equal to the prefix cyclic square of~$r$.
Moreover, since $\sigma$ is $k$-gapped, $u'$ is contained in $w[{\rm end} (v)+1..{\rm end} (v)+k|v|]$.
Therefore, by Corollary~\ref{numcycsqr} the number of different factors satisfying the above conditions
required for~$u'$ is not greater than
$$
\frac{1}{p}|w[{\rm end} (v)+1..{\rm end} (v)+k|v|]|=\frac{1}{p}k|v|=\frac{1}{p}k |r|=k e(r).
$$
Thus $|P_1(r)|\le k e(r)$.

Now let $\sigma\in P_2(r)$. Denote again by $u$ ($u'$) the prefix of length $2p$ in $v$ ($v'$).
Note that in this case $v'$ is determined as $w[{\rm beg} (u')..{\rm end} (r')]$ where $r'$
is determined as the extension of~$u'$. Thus $\sigma$ is determined uniquely by the factor~$u'$.
As in the case~1, we have that $u'$ is a primitive square with period~$p$ which is equal to 
the prefix cyclic square of~$r$. Moreover, since $\sigma$ is $k$-gapped and,
according to Lemma~\ref{overlap}, $r'$ can not overlap with~$r$ by at least $p$ letters,
$u'$ is contained in the factor $w[{\rm end} (r)+1-p..{\rm end} (r)+k|v|]$ which is
contained in $w[{\rm end} (r)+1-p..{\rm end} (r)+k|r|]$. Therefore, by Corollary~\ref{numcycsqr} 
the number of different factors satisfying the conditions required for~$u'$ is not greater than
$$
\frac{1}{p}|w[{\rm end} (r)+1-p..{\rm end} (r)+k|r|]|=\frac{1}{p}(k |r|+p)=k e(r)+1.
$$
Thus $|P_2(r)|\le k e(r)+1$.

Finally let $\sigma\in P_3(r)$. Denote by $u$ and $u'$ the suffixes of length $2p$
in $v$ and $v'$ respectively. Note that in this case $v'$ is determined as 
$w[{\rm beg} (r')..{\rm end} (u')]$ where $r'$ is determined as the extension of~$u'$.
Thus $\sigma$ is determined uniquely by the factor~$u'$. Since $u=u'$ and $u$ 
is the suffix cyclic square of~$r$, the factor $u'$ is a primitive square with 
period~$p$ which is equal to the suffix cyclic square of~$r$. Moreover, since 
$\sigma$ is $k$-gapped, $u'$ is contained in the factor $w[{\rm end} (r)+1..{\rm end} (r)+k|v|]$
which is contained in $w[{\rm end} (r)+1..{\rm end} (r)+k|r|]$. Therefore, as in the case~1, 
we obtain that the number of different factors satisfying the conditions required for~$u'$ is not 
greater than $k e(r)$. Thus $|P_3(r)|\le k e(r)$.

Summing up the obtained bounds for $|P_1(r)|$, $|P_2(r)|$, and $|P_3(r)|$, we conclude that 
$|P(r)|\le 3k e(r)+1$.

Since any non-private maximal gapped repeat is generated by some maximal repetition,
Lemma~\ref{numnonpriv} implies immediately that the number of non-private maximal $k$-gapped 
repeats in~$w$ is $O(k{\rm E}(w))$. Therefore, from Theorem~\ref{sumexp} we derive 

\begin{corollary}
The number of non-private maximal $k$-gapped repeats in~$w$ is $O(kn)$.
\label{nonpriv}
\end{corollary}

From Corollaries~\ref{numpriv} and~\ref{nonpriv} we have
\begin{corollary}
$|{\cal PP}_k|=O(kn)$.
\label{onPP}
\end{corollary}

To estimate the number of semiperiodic maximal $k$-gapped repeats in~$w$, we
estimate separately the numbers of prefix semiperiodic and suffix semiperiodic
maximal $k$-gapped repeats in~$w$. Let $\sigma\equiv (v', v'')$ be an arbitrary 
maximal repeat from ${\cal PSP}_k$, and $p$ be the minimal period of 
periodic prefixes of~$\sigma$. Denote by $u'$ ($u''$) the periodic prefix 
of $v'$ ($v''$), and by $r'$ ($r''$) the extension of $r'$ ($r''$) in~$w$. 
Note that $r'$ and $r''$ are maximal repetitions of~$w$ with the minimal period~$p$. 
From $v'[|u'|+1]=v''[|u''|+1]\neq v'[|u'|+1-p]=v''[|u''|+1-p]$ we have 
$w[{\rm end} (u')+1]\neq w[{\rm end} (u')+1-p]$ and $w[{\rm end} (u'')+1]\neq 
w[{\rm end} (u'')+1-p]$, so
\begin{equation}
{\rm end} (r')={\rm end} (u'),\quad {\rm end} (r'')={\rm end} (u'').
\label{endpref}
\end{equation}
Thus $r'$ and $r''$ are different maximal repetitions in~$w$.
If $|r'|\le |r''|$, we will say that $\sigma$ is {\it generated from the left} 
by the repetition $r'$ per the repetition $r''$. Otherwise we will say that $\sigma$ 
is {\it generated from the right} by the repetition $r''$ per the repetition $r'$. 
A maximal repeat $\sigma$ from~${\cal PSP}_k$ is {\it generated} by a repetition~$r$ 
if $\sigma$ is generated from the left or from the right by~$r$.

\begin{proposition}
If a maximal repeat from~${\cal PSP}_k$ is generated by a repetition~$r$
then $r$ coincides with the periodic prefix of this repeat contained in~$r$.
\label{proponPSP}
\end{proposition}

{\bf Proof.} Assume that the maximal repeat $\sigma$ from~${\cal PSP}_k$
is generated from the left by the repetition~$r'$ (the case when $\sigma$
is generated from the right by the repetition~$r''$ is considered analogously).
According to~(\ref{endpref}), we have that ${\rm end} (r')={\rm end} (u')$.
Let ${\rm beg} (r')<{\rm beg} (u')$. Then from relations~(\ref{endpref}),
$u'=u''$, and $|r'|\le |r''|$ we obtain that ${\rm beg} (r'')<{\rm beg} (u'')$.
So $w[{\rm beg} (u')-1]=w[{\rm beg} (u')-1+p]$ and $w[{\rm beg} (u'')-1]=w[{\rm beg} (u'')-1+p]$.
Since $u'=u''$ we have also that $w[{\rm beg} (u')-1+p]=u'[p]=u''[p]=w[{\rm beg} (u'')-1+p]$.
Thus $w[{\rm beg} (u')-1]=w[{\rm beg} (u'')-1]$, which contradicts that the repeat~$\sigma$ is maximal. 
Hence ${\rm beg} (r')={\rm beg} (u')$, i.e. $r'\equiv u'$.

\begin{proposition}
For any maximal repetitions $r'$, $r''$ in~$w$, at most one maximal repeat from~${\cal PSP}_k$
can be generated from the left by~$r'$ per~$r''$.
\label{proponPSP1}
\end{proposition}

{\bf Proof.} Let $\sigma\equiv (v', v'')$ be an arbitrary maximal repeat from ${\cal PSP}_k$
generated from the left by a repetition~$r'$ per a repetition~$r''$. Then,
using relations~(\ref{endpref}) and Proposition~\ref{proponPSP}, we obtain that 
${\rm beg} (v')={\rm beg} (u')= {\rm beg} (r')$ and ${\rm beg} (v'')={\rm end} (r'')-|u''|+1=
{\rm end} (r'')-|r'|+1$. Denote by $x$ the suffix of $v'$ and $v''$ such that
$v'=u'x=u''x=v''$. Using relations~(\ref{endpref}) and taking into account that
the repeat~$\sigma$ is maximal, it is easy to see that ${\rm end} (v')={\rm end} (r')+|x|$,
${\rm end} (v'')={\rm end} (r'')+|x|$, and $x$ is the greatest common prefix
of $w[{\rm end} (r')+1..n$ and $w[{\rm end} (r'')+1..n$. Thus the copies $v'$ and $v''$
of the repeat~$\sigma$ are uniquely defined by the repetitions $r'$ and $r''$ 
which implies Proposition~\ref{proponPSP1}.

If some maximal repeat from~${\cal PSP}_k$ is generated from the left by a repetition~$r'$
per a repetition~$r''$, we call the repetition~$r''$ {\it left associated} with the repetition~$r'$.

\begin{proposition}
If a repetition~$r''$ is left associated with a repetition~$r'$ then
${\rm end} (r')<{\rm end} (r'')\le {\rm end} (r')+2k|r'|$.
\label{proponPSP2}
\end{proposition}

{\bf Proof.} Let some maximal repeat $\sigma\equiv (v', v'')$ from~${\cal PSP}_k$ be generated 
from the left by the repetition~$r'$ per the repetition~$r''$. It follows from
relations~(\ref{endpref}) that ${\rm end} (r'')-{\rm end} (r')$ is the period of~$\sigma$.
Therefore, since $\sigma$ is $k$-gapped,
$$
0<{\rm end} (r'')-{\rm end} (r')\le k|v'|\le 2k|u'|\le 2k|r'|.
$$
These inequalities imply Proposition~\ref{proponPSP2}.

\begin{lemma}
For any maximal repetition~$r$ in~$w$ there exist no more than $4k$ repetitions
left associated with~$r$.
\label{lemonPSP}
\end{lemma}

{\bf Proof.} Let $p$ be the minimal period of~$r$, and $r_1, r_2,\ldots , r_s$
be all repetitions in~$w$ which are left associated with~$r$ and sorted in non-decreasing 
order of their end positions, i.e. ${\rm end} (r_1)\le {\rm end} (r_2)\le\ldots\le {\rm end} (r_s)$.
Recall that according to the definition of left associated repetitions all the repetitions
$r_1, r_2,\ldots , r_s$ are maximal repetitions with the minimal period~$p$ which are not
shorter than~$r$. So, by Lemma~\ref{overlap}, the overlap of each adjacent 
repetitions $r_{i-1}$ and $r_i$ is less than~$p$. Therefore
$$
{\rm end} (r_i)-{\rm end} (r_{i-1})>|r_i|-p\ge |r|-p\ge |r|/2.
$$
Thus, taking into account Proposition~\ref{proponPSP2}, we have
$$
{\rm end} (r)<{\rm end} (r_1)<{\rm end} (r_2)<\ldots <{\rm end} (r_s)\le {\rm end} (r)+2k|r|
$$
where ${\rm end} (r_i)>{\rm end} (r_{i-1})+|r|/2$. These inequalities
imply that $s<\frac{2k|r|}{|r|/2}+1=4k+1$, i.e. $s\le 4k$.

From Lemma~ref{lemonPSP} and Proposition~\ref{proponPSP1} we immeditely obtain
that for any maximal repetition~$r$ in~$w$ there exist no more than $4k$ repeats 
from~${\cal PSP}_k$ which are generated from the left by~$r$. In the symmetrical way
we can prove that for any maximal repetition~$r$ in~$w$ there exist no more than $4k$
repeats from~${\cal PSP}_k$ which are generated from the right by~$r$. Thus, any
maximal repetition in~$w$ can generate no more tnan $8k$ repeats from~${\cal PSP}_k$.
Therefore, since any repeat from~${\cal PSP}_k$ is generated by some maximal repetition 
in~$w$, from Corollary~\ref{onmaxrun} we obtain

\begin{corollary}
$|{\cal PSP}_k|=O(kn)$.
\label{onPSP}
\end{corollary}

In an analogous way we can prove that $|{\cal SSP}_k|=O(kn)$. Thus we have

\begin{corollary}
$|{\cal SP}_k|=O(kn)$.
\label{onSP}
\end{corollary}

For estimating the number of ordinary maximal $k$-gapped repeats in~$w$ we use the idea which
was used before in~\cite{Kolpakov12}. Namely, we consider pairs of positive integers $(j, p)$. 
We call such pairs {\it points}. For any two points $(j, p')$, $(j'', p'')$ we say that 
the point $(j', p')$ {\it covers} the point $(j'', p'')$ if $p'\le p'' \le p'-\frac{p'}{4k}$ 
and $j'-\frac{p'}{3k}\le j''\le j'$. Let ${\cal Q}$ be the set of all points $(j, p)$ 
such that $1\le j,p\le n$. We represent any maximal repeat~$\sigma$ from ${\cal OP}_k$ by the
point $(j, p)$ in ${\cal Q}$ where $j$ is the end position of the left copy of~$\sigma$
and $p$ is the period of~$\sigma$. It is obvious that $\sigma$ is uniquely defined by the values 
$j$ and~$p$, so two different repeats from ${\cal OP}_k$ can not be represented by the
same point. A point is {\it covered} by~$\sigma$ if the point is covered by the point 
representing~$\sigma$. By $V[\sigma ]$ we denote the set of all points covered by the repeat~$\sigma$. 
We show that any point from ${\cal Q}$ can not be covered by two different repeats 
from ${\cal OP}_k$.

\begin{lemma}
Two different repeats from ${\cal OP}_k$ can not cover the same point.
\label{keylemma}
\end{lemma}

{\bf Proof.} 
Let $\sigma'$, $\sigma''$ be two different repeats from ${\cal OP}_k$ covering the same point $(j, p)$. 
Let $v'\equiv w[i'..j']$ ($v''\equiv w[i''..j'']$) be the left copy of $\sigma'$ ($\sigma''$), and 
$p'$ ($p''$) be the period of $\sigma'$ ($\sigma''$). Thus $\sigma'$ and $\sigma''$ are represented respectively
in ${\cal Q}$ by the points $(j', p')$ and $(j'', p'')$.

First we consider the case $p'=p''$. In this case we obviously have $j'\neq j''$, and without
loss of generality we assume that $j'<j''$. From inequalities $j''-\frac{p''}{3k}\le j\le j'<j''$
and $|v''|\ge\frac{p''}{k}$ we obtain that the letter $w[j'+1]$ is contained in~$v''$. Hence
$w[j'+1]=w[j'+1+p'']=w[j'+1+p']$ which contradicts the maximality of~$\sigma'$. Thus the case $p'=p''$
is impossible.

Now consider the case $p'\neq p''$. Without loss of generality we assume that $p'>p''$.
Define $\delta=p'-p''>0$. From the inequalities $p'-\frac{p'}{4k}\le p\le p''<p'$ we
obtain that $\delta<\frac{p'}{4k}$. To prove that this case is also impossible, we show
that in this case either $\sigma'$ or $\sigma''$ has to be periodic or semi-periodic which contradicts
that both $\sigma'$ and $\sigma''$ are ordinary. We consider separately the following four subcases.

Subcase 1. Let $j''\ge j'$ and $i''\ge i'$. Denote by $u$ the overlap $w[i''..j']$ of the 
factors $v'$ and $v''$. From the inequalities $j''-\frac{p''}{3k}\le j\le j'\le j''$ we have
$j''-j'\le \frac{p''}{3k}\le \frac{|v''|}{3}$, so 
$$
|u|=|v''|-(j''-j')\ge \frac{2|v''|}{3}\ge \frac{2p''}{3k}. 
$$
Since $p'-\frac{p'}{4k}\le p\le p''$, we have also $p''\ge p'\frac{4k-1}{4k}$. Thus
$$
|u|\ge p'\frac{4k-1}{4k}\cdot\frac{2}{3k}>\frac{p'}{2k}>2\delta.
$$
Sinse $u$ is contained in the both left copies $v'$ and $v''$, we obtain that
$u=w[i''+p''..j'+p'']=w[i''+p'..j'+p']$. Thus, $\delta$ is a period of~$u$ and 
$|u|>2\delta$, i.e. $u$ is a repetition. Therefore, since $|u|\ge \frac{2|v''|}{3}>\frac{|v''|}{2}$,
we conclude that in this case $\sigma''$ has to be semi-periodic or periodic.

Subcase 2. Let $j''\ge j'$ and $i''<i'$, i.e. $v'$ is contained in $v''$. Therefore,
$v'=w[i'+p''..j'+p'']=w[i'+p'..j'+p']$, so $\delta$ is a period of~$v'$. Note also
that $|v'|\ge\frac{p'}{k}>4\delta$. Thus $v'$ is a repetition, so $\sigma'$ is periodic
in this case.

Subcase 3. Let $j''<j'$ and $i''\ge i'$, i.e. $v''$ is contained in $v'$. Therefore,
$v'=w[i''+p''..j''+p'']=w[i''+p'..j''+p']$, so $\delta$ is a period of~$v''$. Note also
that $p''\ge p'\frac{4k-1}{4k}>\frac{3}{4}p'$, so $v''\ge \frac{p''}{k}>\frac{3}{4k}p'>3\delta$.
Thus $v''$ is a repetition, so $\sigma''$ is periodic in this case.

Subcase 4. Let $j''<j'$ and $i''<i'$. Denote by $u$ the overlap $w[i'..j'']$ of the 
factors $v'$ and $v''$. From the inequalities $j'-\frac{p'}{3}\le j\le j''<j'$ we have
$j'-j''\le \frac{p'}{3k}\le \frac{|v'|}{3}$, so
$$
|u|=|v'|-(j'-j'')\ge \frac{2|v'|}{3}\ge \frac{2p'}{3k}>\frac{8}{3}\delta.
$$
Sinse $u$ is contained in the both left copies $v'$ and $v''$, we obtain that
$u=w[i'+p''..j''+p'']=w[i'+p'..j''+p']$. Thus, $\delta$ is a period of~$u$ and 
$|u|>2\delta$, i.e. $u$ is a repetition. Therefore, since $|u|\ge \frac{2|v'|}{3}>\frac{|v'|}{2}$,
we conclude that in this case $\sigma'$ has to be semi-periodic or periodic.

From Lemma~\ref{keylemma} we obtain

\begin{lemma}
$|{\cal OP}_k|=O(nk^2)$.
\label{OPklemma}
\end{lemma}

{\bf Proof.} 
To prove the lemma, we assign to each point $(j, p)$ the weight $\rho (j, p)=1/p^2$, 
and for any finite set~$A$ of points we define 
$$
\rho (A)=\sum_{(j, p)\in A} \rho (j, p)=\sum_{(j, p)\in A}\frac{1}{p^2}.
$$
Let $\sigma$ be an arbitrary repeat from ${\cal OP}_k$. Then
$$
\rho (V[\sigma ])=\sum_{j-\frac{p}{3k}\le i\le j}\bigl(\sum_{p-\frac{p}{4k}\le q\le p}\frac{1}{q^2}\bigr)>
\frac{p}{3k}\sum_{p-\frac{p}{4k}\le q\le p}\frac{1}{q^2}
$$
where $(j, p)$ is the point representing~$\sigma$. For further estimating of $\rho (V[\sigma ])$ we consider
separately the cases $p<4k$ and $p\ge 4k$. Let $p<4k$. Then
$$
\frac{p}{3k}\sum_{p-\frac{p}{4k}\le q\le p}\frac{1}{q^2}=\frac{p}{3k}\cdot\frac{1}{p^2}=
\frac{1}{3kp}>\frac{1}{12k^2}.
$$
Now let $p\ge 4k$. Then
\begin{eqnarray*}
\sum_{p-\frac{p}{4k}\le q\le p}\frac{1}{q^2}&=&\sum_{q=p-\lfloor\frac{p}{4k}\rfloor}^p\frac{1}{q^2}>
\int_{p-\lfloor\frac{p}{4k}\rfloor}^p\frac{1}{x^2}\,dx\\
&=&\frac{1}{p-\lfloor\frac{p}{4k}\rfloor}-\frac{1}{p+1}=
\frac{1+\lfloor\frac{p}{4k}\rfloor}{(p-\lfloor\frac{p}{4k}\rfloor)(p+1)}\\
&>&\frac{p/4k}{(p-1)(p+1)}>\frac{p/4k}{p^2}=\frac{1}{4kp}.
\end{eqnarray*}
Therefore,
$$
\frac{p}{3k}\sum_{p-\frac{p}{4k}\le q\le p}\frac{1}{q^2}>\frac{p}{3k}\cdot\frac{1}{4kp}=\frac{1}{12k^2}.
$$
Thus, for any repeat~$\sigma$ from ${\cal OP}_k$ we have $\rho (V[\sigma ])>\frac{1}{12k^2}$.
Using this estimation, we obtain that
\begin{equation}
\sum_{\sigma\in {\cal OP}_k}\rho (V[\sigma ])>\frac{|{\cal OP}_k|}{12k^2}.
\label{lowestim}
\end{equation}
Note that any point covered by repeats from ${\cal OP}_k$ belongs to ${\cal Q}$.
On the other hand, by Lemma~\ref{keylemma}, each point of ${\cal Q}$ can not be covered 
by two repeats from ${\cal OP}_k$. Therefore,
$$
\sum_{\sigma\in {\cal OP}_k}\rho (V[\sigma ])\le \rho ({\cal Q})=
\sum_{j=1}^n\sum_{p=1}^n\frac{1}{p^2}=n\sum_{p=1}^n\frac{1}{p^2}<
n\sum_{p=1}^{\infty}\frac{1}{p^2}=\frac{n\pi^2}{6}.
$$
Thus, using inequality~(\ref{lowestim}), we can conclude that 
$|{\cal OP}_k|<2\pi^2nk^2$.

Summing up Corollaries~\ref{onPP} and~\ref{onSP} and Lemma~\ref{OPklemma},
we obtain that for any integer $k\ge 2$ the number of maximal $k$-gapped repeats 
in~$w$ is $O(k^2n)$. This upper bound is obviously generalized to the case of maximal 
$\alpha$-gapped repeats for any real~$\alpha >1$. Thus we can conclude

\begin{lemma}
For any $\alpha >1$ the number of maximal $\alpha$-gapped repeats 
in~$w$ is $O(\alpha^2n)$.
\label{onPk}
\end{lemma}

From Lemma~\ref{onPk}, using the Proposition~\ref{relforep},
one can easily derive the following upper bound for maximal $\delta$-subrepetitions.

\begin{corollary}
Let $0<\delta <1$. Then the number of maximal $\delta$-subrepetitions in~$w$ is $O(n/\delta^2)$.
\label{ondelrep}
\end{corollary}

\section{Computing of maximal gapped repeats}

In this section we propose an algorithm for finding of all maximal $\alpha$-gapped repeats
in the given word~$w$ for a fixed value of~$\alpha$. The proposed algorithm is actually
a modification of the algorithm described in~\cite{Lothaire05} for finding all repeats
with a fixed gap in a given word. In particular, the two following basic tools are used
in this modification.

The first tool is special functions which are defined as follows. Let $u$, $v$ be two
arbitrary words. For each $i=2, 3,\ldots , |u|$ we define  ${\rm LP}_{u}(i)$ as the
length of the longest common prefix of $u$ and $u[i..|u|]$. For each $i=1, 2,\ldots , |u|-1$
we define ${\rm LS}_{u}(i)$ as the length of the longest common suffix of $u$ and $u[1..|u|-i]$.
For each $i=0, 1,\ldots , |u|-1$ we define ${\rm LP}_{u|v}(i)$ as the length of the 
longest common prefix of $u[|u|-i..|u|]v$ and $v$. For each  $i=1, 2,\ldots , |v|$
we define ${\rm LS}_{u|v}(i)$ as the length of the longest common suffix of $u$ and
$uv[1..i]$. The functions ${\rm LP}_{u}$ and ${\rm LS}_{u}$ can be computed in $O(|u|)$
time and the functions ${\rm LP}_{u|v}$ and ${\rm LS}_{u|v}$ can be computed in 
$O(|u|+|v|)$ time (see, e.g.,~\cite{Lothaire05}).
 
The second tool is a factorization $f\equiv f_1f_2\ldots f_t$ of the word~$w$ which
is called {\it non-overlapping $s$-factorization} and defined inductively as follows:
\begin{itemize}
\item $f_1\equiv w[1]$.
\item Let for $i>1$ the factors $f_1$, \ldots, $f_{i-1}$ are already computed, and
$w[j]$ be the letter which follows the factor $f_{i-1}$ in~$w$. Then $f_i\equiv w[j]$
if the letter $w[j]$ has no occurences in $f_1f_2\ldots f_{i-1}$; otherwise $f_i$ is
the longest factor in~$w$ which follows~$f_{i-1}$ and has an occurence in $f_1f_2\ldots f_{i-1}$.
\end{itemize}
The factorization~$f$ can be computed in $O(n)$ time for the case of constant alphabet size 
and in $O(n\log n)$ time for the general case (see, e.g.,~\cite{Lothaire05}).
By $a_i$ ($b_i$) we denote the start (end) position of the factor $f_i$. The length of~$f_i$
is denoted by $l_i$. For $i=1, 2,\ldots, t-1$ we will consider also the factor
$w[a_i..b_i+1]$ which is denoted by $f'_i$.

For convenience sake we consider the case when $\alpha$ is integer, i.e. for any integer $k\ge 2$
we describe the algorithm of finding in~$w$ all repeats from ${\cal GR}_k(w)$. To this purpose
we divide the set ${\cal GR}_k(w)$ into the following two nonoverlapping subsets: ${\cal FGR}$ is the
set of all repeats from ${\cal GR}_k(w)$ which are not strictly contained in any factor~$f_i$
of the factorization~$f$, and ${\cal SGR}$ is the set of all repeats from ${\cal GR}_k(w)$ 
which are strictly contained in factors of the factorization~$f$. To compute the set 
${\cal GR}_k(w)$, we compute separately the sets ${\cal FGR}$ and ${\cal SGR}$. For each
$i=2, 3,\ldots , t$ we define in the set ${\cal FGR}$ the following subsets: ${\cal FGR}'_i$
is the set of all repeats $\sigma$ from ${\cal FGR}$ such that
\begin{enumerate}
\item $b_{i-1}<{\rm end}(\sigma)\le b_i$;
\item ${\rm beg}(\sigma)\le a_i$;
\end{enumerate}
and ${\cal FGR}''_i$ is the set of all repeats $\sigma$ from ${\cal FGR}$ such that
\begin{enumerate}
\item ${\rm end}(\sigma)=b_i$;
\item ${\rm beg}(\sigma)> a_i$.
\end{enumerate}
It is easy to see that all the subsets ${\cal FGR}'_i$ and ${\cal FGR}''_i$ are
nonoverlapping. Moreover, taking into account that the factor $f_1$ consists of
only one letter, we have that ${\cal FGR}=\bigcup_{i=2}^t {\cal FGR}'_i\cup
\bigcup_{i=2}^t {\cal FGR}''_i$. To compute the set ${\cal FGR}$, we compute 
separately the sets ${\cal FGR}'_i$ and ${\cal FGR}''_i$ for $i=2, 3,\ldots , t$.

To compute ${\cal FGR}'_i$, we consider in this set the following nonoverlapping 
subsets: ${\cal FGR}^{\rm lrt}_i$ is the set of all repeats from ${\cal FGR}'_i$ 
which left copies contain the frontier between the factors $f_{i-1}$ and $f_i$, 
${\cal FGR}^{\rm rrt}_i$ is the set of all repeats from ${\cal FGR}'_i$ 
which right copies contain the frontier between the factors $f_{i-1}$ and $f_i$, 
and ${\cal FGR}^{\rm mid}_i$ is the set of all repeats $\sigma$ from ${\cal FGR}'_i$ 
such that neither left nor right copies of~$\sigma$ contain the frontier between 
the factors $f_{i-1}$ and $f_i$. It is obvious that
${\cal FGR}'_i={\cal FGR^{\rm lrt}}_i\cup {\cal FGR}^{\rm rrt}_i\cup {\cal FGR}^{\rm mid}_i$.
We compute separately the considered subsets of ${\cal FGR}'_i$.

1. {\bf Computing the set ${\cal FGR}^{\rm lrt}_i$.} Let $\sigma\equiv 
(w[i' .. j'], w[i'' .. j''])$ be a repeat from ${\cal FGR}^{\rm lrt}_i$ with
a period~$p$. Note that in this case $p\le l_i$, so $c(\sigma)<l_i$. 
Thus, $\sigma$ is strictly contained in the factor $w[a_i-l_i .. a_{i+1}]\equiv 
g_if'_i$ where $g_i\equiv w[a_i-l_i .. b_{i-1}]$.
Since $w[i' .. j']$ contains the frontier between $f_{i-1}$ and $f_i$,
we have $i'-1\le b_{i-1}\le j'$. Thus, we can consider the factors
$w[a_i .. j']$ and $w[i' .. b_{i-1}]$. Note that $w[a_i .. j']=w[a_i+p .. j'']$
and $w[j'+1]\neq w[j''+1]$. Moreover, from the condition ${\rm end}(\sigma)\le b_i$
we have $j''\le b_i$. Therefore, $w[a_i .. j']$ is the longest common prefix
of $f'_i$ and $f'_i[p+1 .. l'_i]$, i.e. $|w[a_i .. j']|=|w[a_i+p .. j'']|=
{\rm LP}_{f'_i}(p+1)$. Note also that $w[i' .. b_{i-1}]=w[i'' .. b_{i-1}+p]$
and $w[i'-1]\neq w[i''-1]$. Moreover, $|w[i' .. b_{i-1}]|\le c(\sigma)<l_i$.
Therefore, $w[i' .. b_{i-1}]$ is the longest common suffix of the words 
$g_i$ and $g_if_i[1 .. p]$, i.e. $|w[i' .. b_{i-1}]|=
|w[i'' .. b_{i-1}+p]|={\rm LS}_{g_i|f_i}(p)$. Thus,
\begin{equation}
\sigma\equiv (w[a_i-\hat{\rm LS}(p) .. b_{i-1}+\hat{\rm LP}(p)],
w[a_i+p-\hat{\rm LS}(p) .. b_{i-1}+p+\hat{\rm LP}(p)])
\label{sigmalrt}
\end{equation}
where $\hat{\rm LP}(p)={\rm LP}_{f'_i}(p+1)$ and $\hat{\rm LS}(p)={\rm LS}_{g_i|f_i}(p)$,
i.e. $\sigma$ is defined uniquely by the period~$p$. Since 
$\hat{\rm LP}(p)+\hat{\rm LS}(p)=c(\sigma)$ and $\sigma$ is
a $k$-gapped repeat, we have the following restrictions
for $\hat{\rm LP}(p)$ and $\hat{\rm LS}(p)$:
\begin{equation}
p/k\le \hat{\rm LP}(p)+\hat{\rm LS}(p)<p.
\label{condlrt1}
\end{equation}
Moreover, from the condition ${\rm end}(\sigma)\le b_i$ we have
the restriction
\begin{equation}
\hat{\rm LP}(p)\le l_i-p.
\label{condlrt2}
\end{equation}
On the other hand, if for some~$p$ such that $p\le l_i$ the conditions
(\ref{condlrt1}) and~(\ref{condlrt2}) hold, in the set ${\cal FGR}^{\rm lrt}_i$
there exists the maximal $k$-gapped repeat~(\ref{sigmalrt}) with the period~$p$.
Thus, to compute ${\cal FGR}^{\rm lrt}_i$, for each $p=1, 2,\ldots , l_i$
we compute the values $\hat{\rm LP}(p)$ and $\hat{\rm LS}(p)$ and check
the conditions (\ref{condlrt1}) and~(\ref{condlrt2}). If these conditions
are valid we add the corresponding repeat~(\ref{sigmalrt}) to ${\cal FGR}^{\rm lrt}_i$.
As noted above, all the values $\hat{\rm LP}(p)$ and $\hat{\rm LS}(p)$ can
be computed in $O(|g_i|+|f'_i|)=O(l_i)$ time, and all the conditions (\ref{condlrt1}) 
and~(\ref{condlrt2}) can be checked in $O(l_i)$ time. Thus, the set ${\cal FGR}^{\rm lrt}_i$
can be computed in $O(l_i)$ time.

2. {\bf Computing the set ${\cal FGR}^{\rm rrt}_i$.} Let $\sigma\equiv 
(w[i' .. j'], w[i'' .. j''])$ be a repeat from ${\cal FGR}^{\rm rrt}_i$ with
a period~$p$. Then for $\sigma$ we have the following

\begin{proposition}
The right copy of~$\sigma$ doesn't contain the frontier between the
factors $f_{i-2}$ and $f_{i-1}$.
\label{forrt}
\end{proposition}

{\bf Proof.} Assume that the right copy $w[i'' .. j'']$ contains
the frontier between $f_{i-2}$ and $f_{i-1}$. Then we can consider
the factor $u'\equiv w[a_{i-1} .. j'']$ which is a suffix of the
right copy. Since $u'$ is also a suffix of the right copy of $\sigma$,
in $f_1f_2\ldots f_{i-2}$ there is an occurrence of $u'$. Moreover,
the factor $u'$ immediately follows $f_1f_2\ldots f_{i-2}$ and
$|u'|>|f_{i-1}|$ because of $j''={\rm end}(\sigma)>b_{i-1}$.
This contradicts the definition of the factor $f_{i-1}$.

From Proposition~\ref{forrt} and the condition ${\rm end}(\sigma)\le b_i$
we immediately obtain

\begin{corollary}
$c(\sigma)<l_{i-1}+l_i$.
\label{corrt}
\end{corollary}

Thus, $p\le kc(\sigma)<k(l_{i-1}+l_i)$ and $i''>a_{i-1}$ by Proposition~\ref{forrt}.
Therefore, $i'=i''-p>a_{i-1}-k(l_{i-1}+l_i)$, i.e. $\sigma$ is strictly contained 
in the factor $w[a_{i-1}-k(l_{i-1}+l_i) .. a_{i+1}]\equiv g'_if'_i$ where
$g'_i\equiv w[a_{i-1}-k(l_{i-1}+l_i) .. b_{i-1}]$. Since $w[i'' .. j'']$ contains 
the frontier between $f_{i-1}$ and $f_i$, we can consider the factors
$w[a_i .. j'']$ and $w[i'' .. b_{i-1}]$. Note that $w[a_i .. j'']=w[a_i-p .. j']$
and $w[j'+1]\neq w[j''+1]$. Moreover, $j''\le b_i$. Therefore, $|w[a_i .. j'']|=|w[a_i-p .. j']|=
{\rm LP}_{g'_i|f'_i}(p-1)$. Note also that $w[i'' .. b_{i-1}]=w[i' .. b_{i-1}-p]$ 
and $w[i'-1]\neq w[i''-1]$. Thus $|w[i'' .. b_{i-1}]|=|w[i' .. b_{i-1}-p]|=
{\rm LS}_{g'_i}(p)$. Hence
\begin{equation}
\sigma\equiv (w[a_i-p-\hat{\rm LS}(p) .. b_{i-1}-p+\hat{\rm LP}(p)],
w[a_i-\hat{\rm LS}(p) .. b_{i-1}+\hat{\rm LP}(p)])
\label{sigmarrt}
\end{equation}
where $\hat{\rm LP}(p)={\rm LP}_{g'_i|f'_i}(p-1)$ and
$\hat{\rm LS}(p)={\rm LS}_{g'_i}(p)$, i.e. $\sigma$ is 
defined uniquely by the period~$p$. As in the the case of computing 
${\cal FGR}^{\rm lrt}_i$, we have for the period~$p$ the restristions~(\ref{condlrt1}). 
Moreover, since $b_{i-1}<j''\le b_i$, we have the following additional restriction:
\begin{equation}
0<\hat{\rm LP}(p)\le l_i.
\label{condrrt2}
\end{equation}
On the other hand, if for some~$p$ such that $p<k(l_{i-1}+l_i)$ the conditions
(\ref{condlrt1}) and~(\ref{condrrt2}) hold, in the set ${\cal FGR}^{\rm rrt}_i$
there exists the $k$-gapped repeat~(\ref{sigmarrt}) with the period~$p$.
Thus, to compute ${\cal FGR}^{\rm rrt}_i$, for each $p<k(l_{i-1}+l_i)$
we check the conditions (\ref{condlrt1}) and~(\ref{condrrt2}) for the values 
$\hat{\rm LP}(p)$ and $\hat{\rm LS}(p)$. If these conditions hold we add 
the corresponding repeat~(\ref{sigmarrt}) to ${\cal FGR}^{\rm rrt}_i$. Note that
all the values $\hat{\rm LP}(p)$ and $\hat{\rm LS}(p)$ can be computed in 
$O(|g'_i|+|f'_i|)=O(k(l_{i-1}+l_i))$ time, and all the conditions (\ref{condlrt1}) 
and~(\ref{condrrt2}) can be checked in $O(k(l_{i-1}+l_i))$ time. Thus, the set 
${\cal FGR}^{\rm rrt}_i$ can be computed in $O(k(l_{i-1}+l_i))$ time.

3. {\bf Computing the set ${\cal FGR}^{\rm mid}_i$.} Note that the right copies
of all repeats from ${\cal FGR}^{\rm mid}_i$ are strictly contained in $f'_i$. 
Let $q=\lfloor\log{k/(k-1)}l_i\rfloor$. We denote by $d_s$ the position 
$\lfloor ((k-1)/k)^sl_i\rfloor +1$ for $s= 0, 1,\ldots , q$ and divide the set 
${\cal FGR}^{\rm mid}_i$ into nonoverlapping subsets $MP_1, MP_2,\ldots , MP_q$
where $MP_s$ is the set of all repeats from ${\cal FGR}^{\rm mid}_i$ which
right copies cover the letter $f'_i[d_s]$ but don't cover the letter $f'_i[d_{s-1}]$.

\begin{proposition}
${\cal FGR}^{\rm mid}_i=\bigcup_{s=1}^q MP_s$.
\label{formid}
\end{proposition}

{\bf Proof.} Let $\sigma\equiv (w[i' .. j'], w[i'' .. j''])$ be a repeat from 
${\cal FGR}^{\rm mid}_i$. Since the right copy $w[i'' .. j'']$ doesn't cover
the letter $f'_i[d_0]\equiv w[a_{i+1}]$, for proving the proposition we have
to show that $w[i'' .. j'']$ covers at lest one of the letters $f'_i[d_1], f'_i[d_2],
\ldots , f'_i[d_q]$. Let $w[i'' .. j'']$ do not cover any of these letters.
It is easy to check that $d_q\le 2$, so $w[i'' .. j'']$ can not be to the 
left of the letter $f'_i[d_q]$. Thus, $w[i'' .. j'']$ has to be situated
between some letters $f'_i[d_s]$ and $f'_i[d_{s-1}]$. Then 
\begin{eqnarray*}
c(\sigma )&=&|w[i'' .. j'']|\le d_{s-1}-1-d_s=\left\lfloor\left(\frac{k-1}{k}\right)^{s-1}l_i\right\rfloor
-\left\lfloor \left(\frac{k-1}{k}\right)^sl_i\right\rfloor -1\\
&<&\left(\frac{k-1}{k}\right)^{s-1}l_i-\left(\frac{k-1}{k}\right)^sl_i=
\left(\frac{k-1}{k}\right)^{s-1}\frac{l_i}{k}.
\end{eqnarray*}
Moreover, since the left copy $w[i' .. j']$ is to the left of the letter $w[a_i]$,
we have 
$$
p(\sigma)- c(\sigma)\ge d_s>\left(\frac{k-1}{k}\right)^sl_i>(k-1)c(\sigma).
$$
Thus $p(\sigma)>k c(\sigma)$, which contradicts the assumption that
$\sigma$ is $k$-gapped.

Using Proposition~\ref{formid}, for computing ${\cal FGR}^{\rm mid}_i$
we compute separately the sets $MP_1, MP_2,\ldots , MP_q$. In order to
compute the set $MP_s$, consider an arbitrary repeat $\sigma\equiv 
(w[i' .. j'], w[i'' .. j''])$ with a period~$p$ in this set. Note
that in this case the right copy of~$\sigma$ is strictly contained 
in $f'_i[1 .. d_{s-1}]$, so $c(\sigma)<d_{s-1}$. Thus, $j<k d_{s-1}$
and $\sigma$ is strictly contained in 
$$
w[a_i-k d_{s-1} .. b_{i-1}]f'_i[1 .. d_{s-1}]\equiv
h_{is}h'_{is}
$$
where $h_{is}\equiv w[a_i-k d_{s-1} .. b_{i-1}]f'_i[1 .. d_{s}-1]$
and $h'_{is}\equiv f'_i[d_s .. d_{s-1}]$. Since $w[i'' .. j'']$
covers the letter $f'_[d_s]$ we can consider the factors 
$w[i' .. b_{i-1}+d_s-p]$, $w[a_{i-1}+d_s-p .. j']$,
$w[i'' .. b_{i-1}+d_s]$, $w[a_{i-1}+d_s .. j'']$ and
note that 
$$
|w[i' .. b_{i-1}+d_s-p]|=|w[i'' .. b_{i-1}+d_s]|={\rm LS}_{h_{is}}(p),
$$
$$
|w[a_{i-1}+d_s-p .. j']|=|w[a_{i-1}+d_s .. j'']|={\rm LP}_{h_{is}|h'_{is}}(p-1).
$$
Thus, $\sigma$ is defined uniquely by the period~$p$ as
\begin{equation}
(w[a_i+d_s-p-\hat{\rm LS}(p) .. b_{i-1}+d_s-p+\hat{\rm LP}(p)],
w[a_i+d_s-\hat{\rm LS}(p) .. b_{i-1}+d_s+\hat{\rm LP}(p)])
\label{sigmamid}
\end{equation}
where $\hat{\rm LS}(p)={\rm LS}_{h_{is}}(p)$ and $\hat{\rm LP}(p)={\rm LP}_{h_{is}|h'_{is}}(p-1)$.
Since the repeat~$\sigma$ is $k$-gapped, the conditions~(\ref{condlrt1})
have to be valid for the period~$p$. Moreover, $p$ has to satisfy the
additional restrictions
\begin{equation}
\hat{\rm LP}(p)\le p-d_s,
\label{condmid1}
\end{equation}
\begin{equation}
0<\hat{\rm LP}(p)\le d_{s-1}-d_s,
\label{condmid2}
\end{equation}
\begin{equation}
\hat{\rm LS}(p)<d_s-1,
\label{condmid3}
\end{equation}
following from the definition of the set $MP_s$. On the other hand, for each~$p$ 
satisfying the inequality $p<k d_{s-1}$ and the conditions~(\ref{condlrt1}),
(\ref{condmid1}), (\ref{condmid2}), and~(\ref{condmid3}), there exists the 
$k$-gapped repeat~(\ref{sigmamid}) with the period~$p$ in the set $MP_s$.
Thus, to compute $MP_s$, we check the conditions~(\ref{condlrt1}), (\ref{condmid1}), 
(\ref{condmid2}), and~(\ref{condmid3}) for each~$p$ such that $p<k d_{s-1}$.
If for some~$p$ these conditions hold  we add the corresponding repeat~(\ref{sigmamid}) 
to $MP_s$. Note that the time required for computing the involved values $\hat{\rm LP}(p)$ 
and $\hat{\rm LS}(p)$ is bounded by $O(|h_{is}|+|h'_{is}|)=O(kd_{s-1})$ and the total time
required for checking these conditions is bounded by $O(kd_{s-1})$. Thus, $MP_s$
can be computed in $O(kd_{s-1})=O(((k-1)/k)^{s-1}kl_i)$ time. Hence, taking
into account that
$$
\sum_{s=1}^q\left(\frac{k-1}{k}\right)^{s-1}kl_i<
kl_i\sum_{s=0}^\infty\left(\frac{k-1}{k}\right)^s=k^2l_i,
$$
by Proposition~\ref{formid} we obtain that ${\cal FGR}^{\rm mid}_i$ can be computed in 
$O(k^2l_i)$ time.

Summing up the obtained time bounds for computing the sets ${\cal FGR}^{\rm lrt}_i$,
${\cal FGR}^{\rm rrt}_i$ and ${\cal FGR}^{\rm mid}_i$, we conclude that ${\cal FGR}'_i$
can be computed in $O(kl_{i-1}+k^2l_i)$ time.

It is easy to note that the set ${\cal FGR}''_i$ can also be computed in $O(l_i)$ time
by a simplified version of the described above algorithm for computing ${\cal FGR}^{\rm rrt}_i$.

The set ${\cal SGR}$ is also divided into nonoverlapping subsets ${\cal SGR}_2, {\cal SGR}_3,\ldots ,
{\cal SGR}_t$ where  ${\cal SGR}_i$ is the set of all repeats from ${\cal SGR}$ which 
are strictly contained in $f_i$. These subsets are computed separately by
the procedure described below.

Now we give a general description of the algorithm for computing ${\cal GR}_k(w)$.
Initially we compute the factorization~$f$ for~$w$. During the computation
of~$f$, for each factor $f_i$ such that $|f_i|>1$ we store a pointer to an
occurrence of $f_i$ in $f_1f_2\ldots f_{i-1}$ (such occurence exists by the
definition of non-overlapping $s$-factorization and will be denoted by $v_i$). 
More exactly, we store the difference $\Delta_i$ between the start positions 
of $f_i$ and $v_i$. Computation of values $\Delta_i$ does not affect the time 
complexity of computing the factorization~$f$. Further we execute the following 
procedure of finding all repeats from ${\cal GR}_k(w)$. During this procedure all found 
repeats are stored in lists ${\rm start}[j]$ for $j=1, 2,\ldots , n$ where 
${\rm start}[j]$ is a list of all found repeats with the start position~$j$ sorted 
in non-decreasing order of their end positions. The pocedure consists of $t-1$ 
consecutive steps. At the $i-1$-th step we find all repeats~$\sigma$ from ${\cal GR}_k(w)$ 
such that $a_i\le {\rm end} (\sigma)\le b_i$, i.e. after $i-1$-th step we have found 
all repeats~$\sigma$ from ${\cal GR}_k(w)$ such that ${\rm end} (\sigma)\le b_i$. 
Thus, after the last step all repeats from ${\cal GR}_k(w)$ are found. The $i-1$-th step
is executed as follows. First we compute the set ${\cal FGR}'_i$ as described
above. During the computation all found repeats from ${\cal FGR}'_i$ are initially
stored in auxiliary lists ${\rm fin} [j]$ for $j=a_i, a_i+1,\ldots , b_i$ where
in the list ${\rm fin} [j]$ we store all found repeats with the end position~$j$.
After the computation we process consecutively all lists ${\rm fin} [j]$ in
the increasing order of~$j$ by replacing all found repeats from these lists into
the lists ${\rm start}[j]$ according to their start positions. The auxiliary
sorting through the lists ${\rm fin} [j]$ guarantees that all found repeats will
be placed into the lists ${\rm start}[j]$ in the required order. Further, if
$|f_i|>1$, we compute the set ${\cal SGR}_i$. For computing this set consider an 
arbitrary repeat $\sigma\equiv (u, v)$ from ${\cal SGR}_i$. Since $\sigma$ is 
strictly contained in $f_i$, there exists the occurence 
$$
\sigma'\equiv (w[{\rm beg} (u)-\Delta_i  .. {\rm end} (u)-\Delta_i],
 w[{\rm beg} (v)-\Delta_i  .. {\rm end} (v)-\Delta_i] 
$$
of $\sigma$ which is strictly contained in $v_i$, i.e. ${\rm beg} (\sigma')>{\rm beg} (v')$ 
and ${\rm end} (\sigma')<{\rm end} (v')\le {\rm end} f_{i-1}$. It is obvious that $\sigma'$ 
is also a maximal $k$-gapped repeat from ${\cal GR}_k(w)$, so it has to be found before the $i-1$ 
step. Thus $\sigma'$ is contained in the list ${\rm start}[{\rm beg} (\sigma)-\Delta_i]$. 
On the other hand, for each repeat $\sigma'\equiv (u', v')$ which is contained in a list 
${\rm start}[j]$ where ${\rm beg} (v')<{\rm beg} (\sigma')<{\rm end} (v')$ ans satisfies
the condition ${\rm end} (\sigma')<{\rm end} (v')=b_i-\Delta_i$ there exists the repeat
\begin{equation}
\sigma\equiv (w[{\rm beg} (u')+\Delta_i  .. {\rm end} (u')+\Delta_i],
 w[{\rm beg} (v')+\Delta_i  .. {\rm end} (v')+\Delta_i] 
\label{sigmasec}
\end{equation}
in the set ${\cal SGR}_i$. Thus, to compute ${\cal SGR}_i$ it is enough for any~$j$
such that $a_i<j<b_i$ to copy each repeat $\sigma'\equiv (u', v')$ from the list 
${\rm start}[j-\Delta_i]$ such that ${\rm end} (\sigma')<b_i-\Delta_i$ into
the new list ${\rm start}[j]$ as the repeat $\sigma$ defined in~(\ref{sigmasec})
with preserving the order of repeats in the lists. It can be done in $O(l_i+|{\cal SGR}_i|)$
time, so ${\cal SGR}_i$ can be computed in this time. Finally we compute the set
${\cal FGR}''_i$ in $O(l_i)$ time. During the computation of this set each
found repeat $\sigma$ is placed into the respective list ${\rm start}[{\rm beg} (\sigma)]$.
It is easy to see that at the $i-1$-th step all repeats~$\sigma$ from ${\cal GR}_k(w)$ such that
$a_i\le {\rm end} (\sigma)\le b_i$ will be found and placed into the lists ${\rm start}[j]$ 
in the required order. The time complexity bound for the $i-1$-th step is
$O(kl_{i-1}+k^2l_i+|{\cal SGR}_i|)$. It easily implies $O(k^2n+|{\cal SGR}|)$
total time complexity bound for all steps. Sinse $|{\cal SGR}|<|{\cal GR}_k(w)|=O(k^2n)$
by Lemma~\ref{onPk}, we obtain $O(k^2n)$ time complexity bound for the 
the described procedure of finding all repeats from ${\cal GR}_k(w)$. Taking into
account the time for constructing the factorization~$f$, we conclude that
${\cal GR}_k(w)$ can be computed in $O(k^2n)$ time for the case of constant alphabet size
and in $O(n\log n + k^2n)$ time for the general case. 

For convenience sake we have considered the case of maximal $k$-gapped repeats where
$k$ is integer but it easy to see that the proposed algorithm can be directly generalized
to the case of maximal $\alpha$-gapped repeats for real $\alpha >1$ with
preserving the upper bound for time complexity. Thus we have

\begin{theorem}
For any real $\alpha >1$ all maximal $\alpha$-gapped repeats in~$w$ can be computed in 
$O(\alpha^2n)$ time for the case of constant alphabet size and in $O(n\log n + \alpha^2n)$ 
time for the general case.
\label{fndgapair}
\end{theorem}

Now consider the problem of finding all maximal $\delta$-subrepetitions in a word for a fixed~$\delta$.
Because of the established above one-to-one correspondence between maximal $\delta$-subrepetitions and 
principal $\frac{1}{\delta}$-gapped repeats, this problem is reduced to computing 
all principal $\frac{1}{\delta}$-gapped repeats in a word. We propose the following
algorithm for computing all principal $\frac{1}{\delta}$-gapped repeats in the word~$w$.
Further, for convenience, by the period of a repetition we will mean its minimal period.
First we compute the ordered set ${\cal OSR}_{\delta}$ of all maximal repetitions and all
maximal $\frac{1}{\delta}$-gapped repeats in~$w$ such that all elements of ${\cal OSR}_{\delta}$
are ordered in non-decreasing order of their start positions and, furthermore, 
elements of ${\cal OSR}_{\delta}$ with the same start position are ordered in
increasing order of their periods (it is easy to note that any element of ${\cal OSR}_{\delta}$
determined uniquely by its start position and its period, so the introduced
order in ${\cal OSR}_{\delta}$ is uniquely defined). To compute ${\cal OSR}_{\delta}$,
we find in~$w$ all maximal repetitions and all maximal $\frac{1}{\delta}$-gapped repeats.
Using Theorem~\ref{fndgapair} and the algorithm for finding maximal repetitions proposed
in~\cite{KK00}, it can be done in $O(n/\delta^2)$ time for the case of constant alphabet size 
and in $O(n\log n + n/\delta^2)$ time for the general case. Then we arrange the found
repetitions and repeats in the order required for ${\cal OSR}_{\delta}$. 
By Lemma~\ref{onPk} the number of the maximal $\frac{1}{\delta}$-gapped repeats is
$O(n/\delta^2)$ and by Corollary~\ref{onmaxrun} the number of the maximal repetitions
is $O(n)$, so $|{\cal OSR}_{\delta}|=O(n/\delta^2)$. Therefore, using backet sort, the 
required arrangement can be done in $O(n+|{\cal OSR}_{\delta}|)$ time which is bounded 
by $O(n/\delta^2)$. Thus, ${\cal OSR}_{\delta}$ can be computed in $O(n/\delta^2)$ time 
for the case of constant alphabet size and in $O(n\log n + n/\delta^2)$ time for the 
general case. Note that by Proposition~\ref{princrit2} for discovering all principal repeats 
from the maximal $\frac{1}{\delta}$-gapped repeats it is enough to compute all
stretchable $\frac{1}{\delta}$-gapped repeats in~$w$. To compute stretchable $\frac{1}{\delta}$-gapped 
repeats, we maintain an auxiliary two-way queue ${\rm SRQ}$ consisting of 
elements from ${\cal OSR}_{\delta}$. Elements from ${\cal OSR}_{\delta}$ are
presented by pairs $(p, q)$ where $p$ and $q$ are respectively the period and
the end position of the presented element (it is easy to note that any element of ${\cal OSR}_{\delta}$
determined uniquely by its period and its start position, so two different elements
can not be presented by the same pair in ${\rm SRQ}$). At any time the queue ${\rm SRQ}$
has a form: 
\begin{equation}
(p_1, q_1), (p_2, q_2),\ldots , (p_s, q_s)
\label{queue}
\end{equation}
where $p_1<p_2<\ldots <p_s$ and $q_1<q_2<\ldots <q_s$. Starting from empty ${\rm SRQ}$, 
we try to insert in ${\rm SRQ}$ each element of ${\cal OSR}_{\delta}$ in the prescribed 
order by the following way. The first element of ${\cal OSR}_{\delta}$ is simply inserted
in empty ${\rm SRQ}$. Let an element~$\tau$ with period~$p$ and end position~$q$ be the next 
candidate for insertion in the queue ${\rm SRQ}$ presented in~(\ref{queue}). Firstly we find
the periods $p_i$ and $p_{i+1}$ such that $p_i\le p<p_{i+1}$ and\footnote{We describe our
algorithm for the general case when both $p_i$ and $p_{i+1}$ are exist. The cases when eigther $p_i$
or $p_{i+1}$ does not exist are easily derived from this general case.} compare $q$ with $q_i$.
If $q\le q_i$ we establish that $\tau$ is a stretchable repeat\footnote{It is easy to check
that in this case $\tau$ can not be a repetition.} and don't insert $\tau$ in ${\rm SRQ}$.
Othervise we insert $\tau$ in ${\rm SRQ}$ and remove from ${\rm SRQ}$ all pairs $(p_j, q_j)$
such that $j>i$ and $q_j\le q$ in order to preserve ${\rm SRQ}$ in the proper form. Using
Proposition~\ref{strcrit}, one can check that the described procedure compute correctly
all stretchable repeats from ${\cal OSR}_{\delta}$ which allows to compute all principal 
$\frac{1}{\delta}$-gapped repeats in~$w$. For effective execution of operations required
in this procedure we use the data structure proposed in~\cite{BoasKaasZul}. This data structure
can be constructed in $O(n\log\log n)$ time and allows to execute the operations of finding~$p_i$, 
inserting an element to ${\rm SRQ}$ and removing an element from ${\rm SRQ}$ in $O(\log\log n)$ time.
Note that in the described procedure no more than one of each of these three operations is required 
for treating any element from ${\cal OSR}_{\delta}$. Thus, the time required for computing
all stretchable repeats in ${\cal OSR}_{\delta}$ is $O(n\log\log n + |{\cal OSR}_{\delta})|\log\log n)$,
so can be bounded by $O(n\log\log n/\delta^2)$. Summing up this time bound with the time bound for 
computing the set ${\cal OSR}_{\delta}$, we obtain

\begin{theorem}
Let $0<\delta <1$. Then all maximal $\delta$-subrepetitions in~$w$ can be computed
in $O(\frac{n\log\log n}{\delta^2})$ time for the case of constant alphabet 
size and in $O(n\log n +\frac{n\log\log n}{\delta^2})$ time for the general case.
\label{fndsubrep1}
\end{theorem}

Another algorithm for computing all principal $\frac{1}{\delta}$-gapped repeats in a word
is based on Proposition~\ref{princrit1}. By this proposition, in order to check if a maximal
gapped repeat~$\sigma$ in~$w$ is principal we can compute the minimal period of 
$w[{\rm beg} (\sigma)..{\rm end} (\sigma)]$ and compare this period with $p(\sigma)$: 
if these periods are equal then $\sigma$ is principal; otherwise $\sigma$ is not principal.
The problem of effective answering to queries related to minimal periods of factors
in a word is studied in~\cite{Kociumaketal}. In particular, in~\cite{Kociumaketal} 
a hash table data structure is proposed for resolving this problem. This data structure 
can be constructed in $O(n\log n)$ expected time and allows to compute the minimal period~$p$
of a required factor~$u$ in $O(\log (1+\frac{|u|}{|u|-p}))$ time. Note that for any
$\frac{1}{\delta}$-gapped repeat~$\sigma$ in~$w$ we have $p(w[{\rm beg} (\sigma)..{\rm end} (\sigma)])\le
p(\sigma)\le\frac{|\sigma |}{1+\delta}$, so $p(w[{\rm beg} (\sigma)..{\rm end} (\sigma)])$ 
can be computed in $O(\log (1+\frac{1}{\delta}))$ time. Therefore, using the data structure
from~\cite{Kociumaketal}, for any maximal $\frac{1}{\delta}$-gapped repeat~$\sigma$ in~$w$
we can check if $\sigma$ is principal in $O(\log (1+\frac{1}{\delta}))$ time. Thus,
in our second algorithm we compute the set ${\cal GR}_{1/\delta}(w)$ and for each
repeat~$\sigma$ from~${\cal GR}_{1/\delta}(w)$ check, as described above, if $\sigma$ 
is principal. By Theorem~\ref{fndgapair} the set ${\cal GR}_{1/\delta}(w)$ can be computed 
in $O(n\log n + \frac{n}{\delta^2})$ time. The expected total time for checking all repeats 
from~${\cal GR}_{1/\delta}(w)$ is $O(n\log n+|{\cal GR}_{1/\delta}(w)|\log (1+\frac{1}{\delta}))$, 
so this time can be bounded by $O(n\log n+\frac{n}{\delta^2}\log \frac{1}{\delta})$ since 
$|{\cal GR}_{1/\delta}(w)|=O(\frac{n}{\delta^2})$ by Lemma~\ref{onPk}. Thus we have

\begin{theorem}
Let $0<\delta <1$. Then all maximal $\delta$-subrepetitions in~$w$ can be computed
in $O(n\log n+\frac{n}{\delta^2}\log \frac{1}{\delta})$ expected time.
\label{fndsubrep2}
\end{theorem}

\section{Conclusion}

One of our results is the $O(\alpha^2n)$ upper bound on the number of maximal $\alpha$-gapped 
repeats in a word of length~$n$. On the other hand, it is easy to see that this number can be 
at least\footnote{We will naturally assume that $\alpha\le n$ and $\delta\ge 1/n$.} 
$\Omega (\alpha n)$, so we have a gap between upper and lower bounds on this number.  
Thus we have an open question on the optimality of the obtained upper bound. The performed 
computer experiments show that the order of growth for the maximal number of maximal $\alpha$-gapped 
repeats in a word of length~$n$ is $\alpha n$. It would imply that the order of growth for the maximal 
number of maximal $\delta$-subrepetitions in a word of length~$n$ is $O(n/\delta)$. Checking this 
conjecture would be of interest to us. We assume also that the proposed algorithms are not time optimal, 
so improving these algorithms is another direction for further research.

\section*{Acknowledgments}
 
This work is partially supported by Russian Foundation for Fundamental Research 
(Grant 11-01-00508).


\begin{thebibliography}{10}

\bibitem{Brodal00}
G.~Brodal, R.~Lyngso, C.~Pedersen, J.~Stoye, Finding Maximal Pairs with Bounded Gap,
{\it Journal of Discrete Algorithms}  {\bf 1(1)} (2000), 77--104.

\bibitem{Crochemor81}
M.~Crochemore, An optimal algorithm for computing the repetitions in a word, 
{\it Information Processing Letters} {\bf 12} (1981), 244--250.

\bibitem{CrochRytter95}
M.~Crochemore, W.~Rytter, Squares, cubes, and time-space
efficient string searching, {\it Algorithmica} {\bf 13} (1995), 405--425. 

\bibitem{algonstr}
M.~Crochemore, C.~Hancart, T.~Lecroq, Algorithms on Strings, 
Cambridge University Press, 2007.

\bibitem{CrochIlieTinta}
M.~Crochemore, L.~Ilie, and L.~Tinta, Towards a solution to the "runs" conjecture,
{\it Lecture Notes in Comput. Sci.}  {\bf 5029} (2008), 290--302.

\bibitem{Crochetal1}
M.~Crochemore, C.~Iliopoulos, M.~Kubica, J.~Radoszewski, W.~Rytter, T.~Walen,
Extracting powers and periods in a string from its runs structure,
{\it Lecture Notes in Comput. Sci.} {\bf 6393} (2010), 258--269.

\bibitem{Crochetal11}
M.~Crochemore, M.~Kubica, J.~Radoszewski, W.~Rytter, T.~Walen,
On the maximal sum of exponents of runs in a string,
{\it Lecture Notes in Comput. Sci.} {\bf 6460} (2011), 10--19.

\bibitem{BoasKaasZul}
P.~van~Emde Boas, R.~Kaas, E.~Zulstra, Design and Implementation of an
Efficient Priority Queue, {\it Mathematical Systems Theory} {\bf 10} (1977), 99-127.

\bibitem{GaliSeiferas83}
Z.~Galil, J.~Seiferas, Time-space optimal string matching,
{\it Journal of Computer and System Sciences} {\bf  26(3)} (1983), 280--294.

\bibitem{Gusfield97}
D.~Gusfield, Algorithms on Strings, Trees, and Sequences,
Cambridge University Press, 1997.

\bibitem{GusfStoye04}
D.~Gusfield, J.~Stoye, Linear time algorithms for finding and representing 
all the tandem repeats in a string, {\it Journal of Computer and System Sciences}
{\bf  69(4)} (2004), 525--546.

\bibitem{Kociumaketal}
T.~Kociumaka, J.~Radoszewski, W.~Rytter, T.~Walen, Efficient Data Structures for 
the Factor Periodicity Problem, {\it Lecture Notes in Comput. Sci.} {\bf 7608} (2012), 
284-294.

\bibitem{KK00}
R.~Kolpakov, G.~Kucherov, On Maximal Repetitions in Words, 
{\it Journal of Discrete Algorithms}  {\bf 1(1)} (2000), 159--186.

\bibitem{KK00a}
R.~Kolpakov, G.~Kucherov, Finding Repeats with Fixed Gap,
{\it Proceedings of 7th International Symposium on String
 Processing and Information Retrieval (SPIRE'00)} (2000), 162--168.

\bibitem{Lothaire05}
R.~Kolpakov, G.~Kucherov, Periodic structures in words, chapter for 
{\it the 3rd Lothaire volume Applied Combinatorics on Words}, Cambridge 
University Press, 2005.

\bibitem{KKOch}
R.~Kolpakov, G.~Kucherov, P.~Ochem, On maximal repetitions of arbitrary exponent,
{\it Information Processing Letters}, {\bf 110(7)} (2010), 252--256.

\bibitem{Kolpakov12}
R.~Kolpakov, On primary and secondary repetitions in words,
{\it Theoretical Computer Science}, {\bf 418} (2012), 71--81.

\bibitem{Lothaire83}
M.~Lothaire, Combinatorics on Words, volume~17 of {\it Encyclopedia of
Mathematics and Its Applications}, Addison Wesley, 1983.

\bibitem{Storer88}
J.~Storer, Data compression: methods and theory,
Computer Science Press, Rockville, MD, 1988.


\end{thebibliography}
\end{document}